\documentclass[12pt,preprint]{aastex}
\slugcomment{Submitted to ApJ}

\shorttitle{Effect of Rotation on SASI}
\shortauthors{Yamasaki and Foglizzo}

\begin{document}

\title{Effect of Rotation on the Stability of a Stalled Cylindrical Shock\\
and its Consequences for Core-Collapse Supernovae}

\author{Tatsuya Yamasaki and Thierry Foglizzo}
\affil{Service d'Astrophysique, DSM/IRFU, UMR AIM,
CEA-CNRS-Univ. Paris 7, Saclay, France}
\email{tatsuya.yamasaki@cea.fr, foglizzo@cea.fr}

\begin{abstract}

A perturbative analysis is used to investigate the effect of rotation
on the instability of a steady accretion shock (SASI) in a simple toy-model, in view of better understanding supernova explosions in which the collapsing core contains angular momentum.
A cylindrical geometry is chosen for the sake of simplicity.
Even when the centrifugal force is very small, rotation can have a strong effect on the non-axisymmetric modes of SASI by increasing the growth rate of the spiral modes rotating in the same direction as the steady flow. Counter-rotating spiral modes are significantly damped, while axisymmetric modes are hardly affected by rotation.
The growth rates of spiral modes have a nearly linear dependence on the specific angular momentum of the flow. 
The fundamental one-armed spiral mode ($m=1$) is favoured for small rotation rates, whereas stronger rotation rates favour the mode $m=2$. 
A WKB analysis of higher harmonics indicates that the efficiency of the advective-acoustic cycles associated to spiral modes is strongly affected by rotation in the same manner as low frequency modes, whereas the purely acoustic cycles are stable. 
These results suggest that the linear phase of SASI in rotating core-collapse supernovae naturally selects a spiral mode rotating in the same direction of the flow, as observed in the 3D numerical simulations of Blondin \& Mezzacappa (2007). This emphasizes the need for a 3D approach of rotating core-collapse, before conclusions on the explosion mechanisms and pulsar kicks can be drawn.

\end{abstract}


\section{Introduction\label{intro}}

Despite extensive studies, the explosion mechanism of core-collapse supernovae is still elusive.
According to the delayed explosion scenario,
the shock is first stalled at a distance of a few hundred kilometers, and then revived after neutrinos diffuse out of the proto neutron star. Unfortunately, numerical simulations suggest that neutrino heating may not be efficient enough, at least in spherical symmetry \citep{lieb05}.

Recent studies have shown that the spherical stalled shock is unstable against non radial perturbations
with a low degree $l=1,2$ even if the flow is convectively stable. This result was demonstrated using axisymmetric numerical simulations \citep{blo03,blo06,sch06,sch08,ohn06} and linear stability analyses \citep{gal05,fog07,yam07}. 
Some numerical simulations have shown that this hydrodynamical instability, often called SASI, may assist
the revival of the shock and trigger a successful explosion, powered either by neutrino heating 
\citep{mar07} or by acoustic waves \citep{bur06}.
Some observed properties of young neutron stars may also be the consequences of SASI, such as their distribution of velocities \citep{sch04, sch06} or their spin \citep{blo07a,blo07b}.

Until now, most studies of SASI have assumed that the unperturbed flow is purely radial and not rotating. Since the angular momentum of massive stars is likely to be large \citep{heg05}, it is desirable to understand how the properties of SASI are affected by rotation.
In this Letter, the effect of rotation on the linear stage of SASI is investigated using a perturbative analysis in order to shed light on one of the surprising results observed by \cite{blo07a} in their 3D numerical simulations: the development of SASI seems to systematically favour a spiral mode rotating in the same direction as the accretion flow. As a consequence of momentum conservation, this mode diminishes and may even reverse the angular momentum acquired by the proto-neutron star from the stationary flow. Incidentally, the present linear study does not address another surprising result of \cite{blo07a}, that a spiral mode of SASI always dominate the axisymmetric mode even without rotation.
Following an approach similar to \citet{fog07} (hereafter FGSJ07), we first compute the eigenfrequencies by solving accurately a boundary value problem between the shock surface and the accretor surface; in a second step, we use the same WKB method as in FGSJ07 to measure the stability of purely acoustic and advective-acoustic cycles in this region. This approach is different from \cite{lam07}, which is based on the approximate derivation of a dispersion relation.
Rather than the complexity of describing the non-spherical shape of a shock deformed by rotation \citep{yam05}, we have chosen, as a first step, to solve the much simpler problem of a cylindrical accretion shock. This flow is simple enough to allow for a complete coverage of the parameter space and a physical insight of the main effects of rotation on SASI. Once characterized, these effects can be transposed into the more complex geometry of a rotating stellar core.

\section{Formulation\label{calculations}}

Following a similar desire of simplification, \cite{blo07b} have limited their 2D simulation domain $(r,\varphi)$ to the vicinity of the equatorial region $(\theta=\pi/2)$ of a non-rotating spherical flow, where the poloidal velocity of a symmetric mode is $v_\theta=0$. Their calculation neglected the perturbation of density induced by the term $(\rho/ r)(\partial v_\theta/\partial\theta)$ in the equation of continuity, implicitly decoupling the structure of perturbations in the vertical direction from their structure in the equatorial plane. In order to study the effects of rotation in a simple setup and maintain a self consistent set of equations, we have chosen to adopt a cylindrical geometry invariant along the $z$-axis of rotation. The accretor and the shock are thus cylindrical in our toy-model. 

Except for the geometry of the flow and the presence of rotation, the assumptions are the same as in FGSJ07. In the equations describing the flow (Appendix~A), we have assumed that 
i) the free-falling supersonic flow is cold,
ii) gravity is Newtonian and self-gravity is neglected,
iii) the shock is adiabatic,
iv) neutrino heating is neglected and the cooling rate per volume is approximated as ${\cal L} \propto \rho^{\beta-\alpha}P^{\alpha}$ with $\alpha=3/2$, and $\beta=5/2$, where $\rho$ is the density and $P$ the pressure,
v) the accreting material is described by a perfect gas with a uniform adiabatic index $\gamma=4/3$,
vi) the condition ${\cal M}=0$ is imposed at the inner boundary, where ${\cal M}\equiv -v_r/c$ is the Mach number associated to the radial velocity $v_r$ and adiabatic sound speed $c$,
vii) the flow is inviscid and its specific angular momentum $L$ is conserved.

We have adopted values of the parameters typical for the core-collapse problem: the radius of the inner boundary is $r_*=50$ [km] and the gravitational potential is $\Phi=-1.3GM_{\odot}/r$, where $G$ is the gravitational constant and $M_{\odot}$ the solar mass. By adopting the same gravitational potential as for a spherical accretor, the Bernoulli equation is unchanged.

The stability of the accretion flow is investigated for various values of the specific angular momentum $L$, measured by the rotation frequency $f_{\rm p}\equiv L/(2\pi r_{\rm p}^2)$ extrapolated at a radius $r_{\rm p}\sim 10$[km] by reference to young pulsars. The range of rotation rates considered corresponds to $0\le f_{\rm p}\le 10^{3}$[Hz].
We have chosen to compare instability timescales in cylindrical models with different rotation rates but identical shock radii and mass accretion rates, in order to keep geometrical factors constant. This is made possible by adapting the intensity of cooling accordingly, by a modest amount ($<15\%$) over the range of rotation rates considered. We observe that the advection time from the shock to the accretor is hardly affected by rotation (a few percent). This is because the centrifugal force $L^2/r^3$ is much smaller than gravity $-d\Phi/dr$:
\begin{eqnarray}
{L^2\over r^3 |d\Phi/dr|}&=&4.6\times 10^{-2} \left({r_*\over r}\right)\left({f_{\rm p}\over 10^3{\rm [Hz]}}\right)^2.
\end{eqnarray}
The perturbations superimposed upon the steady solution are proportional to 
$\exp\{-i(\omega t -m\theta -k_z z)\}$.
We adopt the variables $\delta S,\delta q,\delta f,\delta h$ defined by
\begin{eqnarray}
\delta S&\equiv&{2\over\gamma-1}{\delta c\over c}-{\delta\rho\over\rho},\label{eq5}
\\
\delta q&\equiv&\delta\left(\int ^r\frac{\cal L}{\rho v_r}dr'\right),
\label{eq2}\\
\delta f&\equiv& v_r \delta v_r +\frac{L}{r}\delta v_{\theta}+\frac{2}{\gamma -1}c\delta c -\delta q,
\label{eq3}\\
\delta h&\equiv& \frac{\delta v_r}{v_r}+\frac{\delta \rho}{\rho},
\label{eq4}
\end{eqnarray}
where $\delta$ denotes the Eulerian perturbation and $v_\theta$ is the azimuthal velocity.
Then the differential system describing the perturbations becomes particularly compact:
\begin{eqnarray}
\frac{{\rm d} \delta f}{{\rm d} r}&=&\frac{i\omega c^2}{v_r(1-{\cal M}^2)}\left\{{\cal M}^2 \delta h
-{\cal M}^2 \frac{\omega'}{\omega}\frac{\delta f}{c^2}\right.\nonumber\\
&&\left.+[1+(\gamma -1){\cal M}^2 ]\frac{\delta S}{\gamma}-\frac{\delta q}{c^2}\right\},
\label{eq6}\\
\frac{{\rm d} \delta h}{{\rm d} r}&=&\frac{i\omega'}{v_r(1-{\cal M}^2)}\left\{\frac{\mu^2}{c^2}\frac{\omega'}{\omega}\delta f-{\cal M}^2 \delta h
-\delta S +\frac{\delta q}{c^2}\right\},
\label{eq7}\\
\frac{{\rm d}\delta S}{{\rm d} r}&=&\frac{i\omega'}{v_r}\delta S +\delta\left(\frac{\cal L}{Pv_r}\right),
\label{eq8}\\
\frac{{\rm d}\delta q}{{\rm d} r}&=&\frac{i\omega'}{v_r}\delta q +\delta\left(\frac{\cal L}{\rho v_r}\right),
\label{eq9}
\end{eqnarray}
where $\mu$ and $\omega'$ are defined by
\begin{eqnarray}
\mu^2&\equiv& 1-\frac{c^2}{\omega'^2}(1-{\cal M}^2)\left(\frac{m^2}{r^2}+k_z^2 \right),
\label{eq10}\\
\omega'&\equiv&\omega-\frac{mL}{r^2}.
\label{eq11}
\end{eqnarray}
These equations are solved by imposing the Rankine-Hugoniot relations for the perturbed quantities,
which are written as, 
\begin{eqnarray}
\frac{\delta f_{\rm sh}}{\omega}&=&iv_{r,1} \Delta\zeta\left(1-\frac{v_{r,\rm sh}}{v_{r,1}}\right),
\label{eq12}\\
\delta h_{\rm sh}&=&-i\frac{\omega'}{v_{r,\rm sh}}\Delta\zeta\left(1-\frac{v_{r,\rm sh}}{v_{r,1}}\right),
\label{eq13}\\
\frac{\delta S_{\rm sh}}{\gamma}&=&i\frac{\omega' v_{r,1}}{c_{\rm sh}^2}\Delta\zeta\left(1-\frac{v_{r,\rm sh}}{v_{r,1}}\right)^2
-\frac{{\cal L}_{\rm sh}-{\cal L}_1}{\rho_{\rm sh} v_{r,\rm sh}}\frac{\Delta \zeta}{c_{\rm sh}^2}\nonumber\\
&&+\left(1-\frac{v_{r,\rm sh}}{v_{r,1}}\right)\frac{\Delta\zeta}{c_{\rm sh}^2}\left(\frac{v_{r,1} v_{r,\rm sh}}{r_{\rm sh}}+\frac{L^2}{r_{\rm sh}^3}-\frac{{\rm d} \Phi}{{\rm d} r}\right),
\label{eq14}\\
\delta q_{\rm sh}&=&-\frac{{\cal L}_{\rm sh}-{\cal L}_1}{\rho_{\rm sh} v_{r,\rm sh}}\Delta\zeta,
\label{eq15}
\end{eqnarray}
where the subscripts '${\rm sh}$' and '1' refer to the values just below and above the shock, respectively; $\Delta \zeta$ is the radial displacement of the shock surface.
Since we have assumed that the flow above the shock is cold, the cooling rate ${\cal L}_1=0$.
In addition to these equations, we impose the condition that the radial velocity perturbation vanishes
at the inner boundary ($\delta v_r =0$). The derivations of the basic equations and the boundary conditions are shown in the Appendix A.

When rotation is neglected ($L=0$, $\omega'=\omega$), we remark the formal resemblance between the above formulation and the formulation of FGSJ07 describing a spherical flow. The only difference is the expression for the parameter $\mu^2$ in Eq.~(\ref{eq10}), where $(m^2+r^2k_z^2)$ replaces $l(l+1)$, and a geometrical factor $2$ in the boundary condition for the entropy perturbation ($v_{r,\rm sh}v_{r,1}$ replaces $2v_{r,\rm sh}v_{r,1}$ in Eq.~\ref{eq14}).

For a small value of the angular momentum $L$, we also remark that the effect of the centrifugal force $L^2 /r^3$ on the stationary flow is quadratic, and so is its effect on the entropy perturbation in Eq.~(\ref{eq14}). The only first order effect of rotation on the differential system satisfied by $\delta f,\delta h,\delta S,\delta q$, is the Doppler shift described by $\omega'$ in Eq.~(\ref{eq11}).

\section{Results\label{results}}

Given the resemblance of formulations, let us first check that the stability properties of the cylindrical flow without rotation resemble those of the spherical flow studied by FGSJ07 despite the different geometry. 
In the spherical problem, the axisymmetric mode $m=0$ and the spiral modes $\pm m$ of a given degree $l$ have exactly the same growth rate (FGSJ07). In the cylindrical flow, the spiral modes $\pm m$ are also degenerate without rotation. A numerical resolution of the eigenfrequencies shows that the instability is dominated by the mode $m=\pm1$ if $r_{\rm sh}/r_*\ge2$, and by a larger $|m|$ for smaller shock radius, exactly as observed in the spherical flow (Fig. 6 of FGSJ07). The growth rate of the 
axisymmetric mode ($m=0$, $k_z>0$) however, is expected to differ from the spiral modes ($m>0$, $k_z=0$) in a cylindrical flow. For example if $r_{\rm sh}/r_*=5$, the most unstable axisymmetric mode is twice as slow as the most unstable spiral mode.

\begin{figure}
\epsscale{1.0}
\plotone{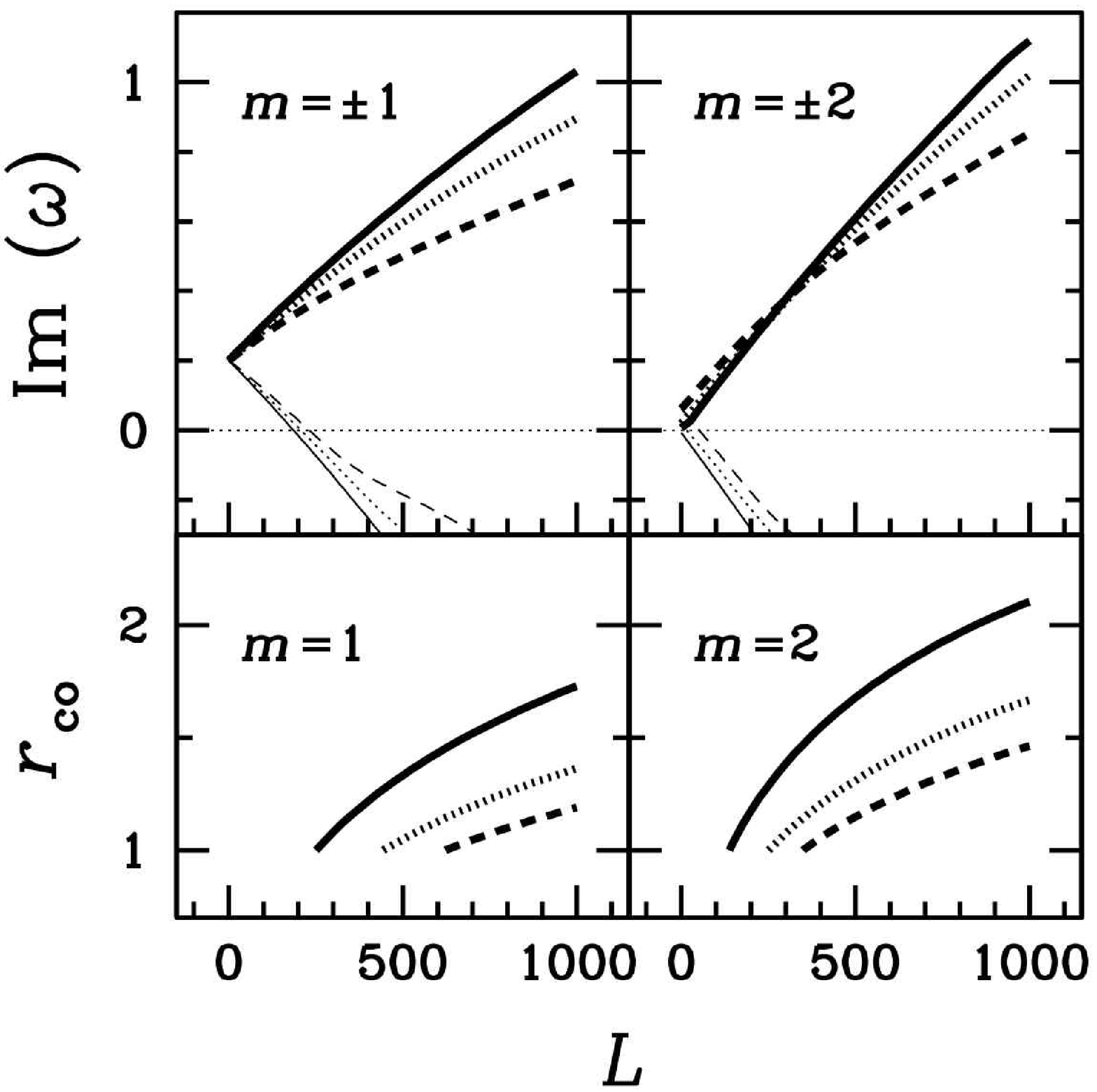}
\caption{Growth rate and corotation radius $r_{\rm co}/r_*$ as a function of the specific angular momentum $L$, when $r_{\rm sh}/r_*=5$. The growth rate is normalized by $|v_{r, \rm sh}|/(r_{\rm sh}-r_*)$. $L$ is normalized by $2\pi\cdot 10^{12}[{\rm cm}^2/{\rm s}]$, and corresponds to the rotation frequency $f_{\rm p}$ [Hz] extrapolated at $10$km. Thick lines are for the modes with $m>0$ and thin curves are for $m<0$. The corotation radii are displayed only for the modes with $m>0$. The solid, dotted and dashed lines are for the fundamental modes, first and second overtones respectively. {\it Left:} The case with $m=\pm 1$. {\it Right:} The case with $m=\pm 2$.
\label{fig1}}
\end{figure}

\begin{figure}
\epsscale{1.0}
\plotone{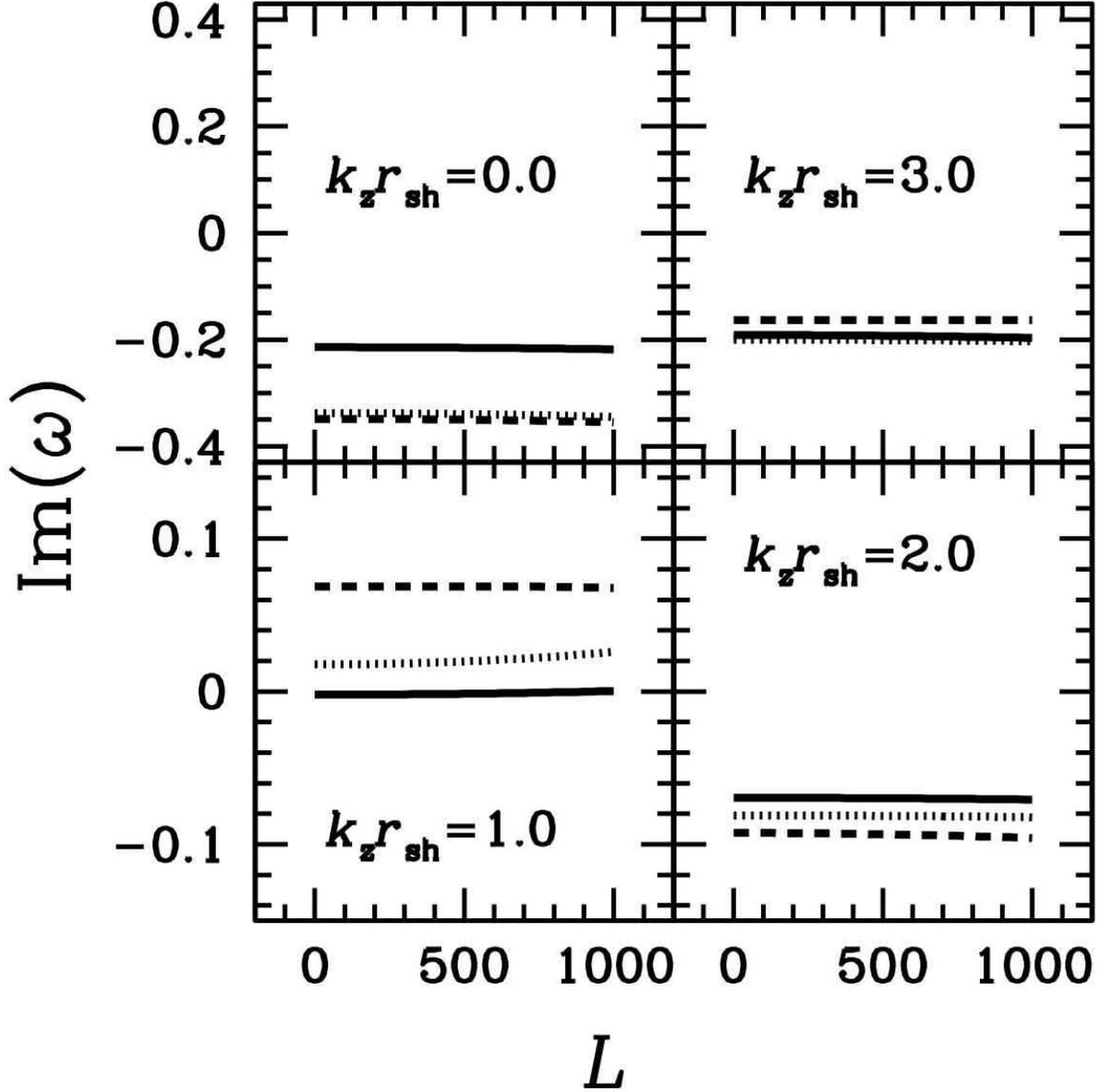}
\caption{Growth rate of the axisymmetric mode ($m=0$) as a function of the specific angular momentum $L$, when $r_{\rm sh}/r_*=5$. The growth rate is normalized by $|v_{r, \rm sh}|/(r_{\rm sh}-r_*)$. $L$ is normalized by $2\pi\cdot 10^{12}[{\rm cm}^2/{\rm s}]$, and corresponds to the rotation frequency $f_{\rm p}$ [Hz] extrapolated at $10$km. The solid, dotted and dashed lines are for the fundamental modes, first and second overtones respectively. The value of the wavenumber $k_z$ is indicated on each plot.\label{fig2}}
\end{figure}

\begin{figure}
\epsscale{1.0}
\plotone{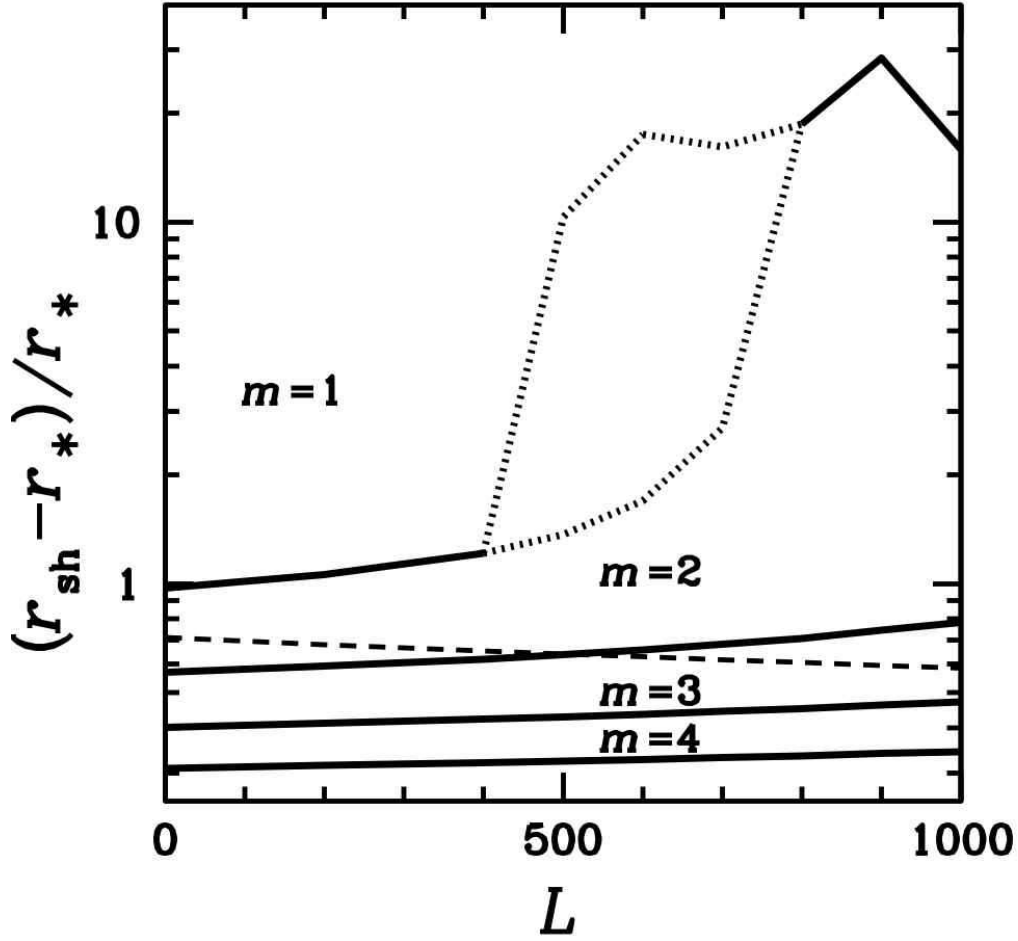}
\caption{Azimuthal wave number $m$ of the mode with the largest growth rate, for each specific angular momentum $L$ and shock radius $r_{\rm sh}$. Units are identical to Fig.~1. In the dotted area, the transition between $m=1$ and $m=2$ is irregular. The spiral mode $m=1$ is unstable in most of the parameter space (above the dashed line).
\label{fig3}}
\end{figure}

The dependence of the growth rate on the specific angular momentum is shown in Fig. \ref{fig1} for the spiral modes $m=\pm1,\pm2$ in a flow where $r_{\rm sh}/r_*=5$.
The growth rate of the modes rotating in the same direction as the flow ($m>0$) is increased by rotation,
whereas the counter-rotating modes ($m<0$) are stabilized. The increase of the growth rate is almost proportional to the specific angular momentum. In a rotating flow with $L\sim 200\times 2\pi\cdot 10^{12}[{\rm cm}^2/{\rm s}]$ ($f_{\rm p}=200[{\rm Hz}]$), the growth rate of the fundamental mode $m=1$ is twice its value in a non rotating flow of same size. 
The strong effect of rotation on the growth rate of SASI does not seem to depend on the presence of a corotation radius $r_{\rm co}$, defined by ${\rm Re}(\omega) -mL/r_{\rm co}^2=0$, also displayed in Fig. \ref{fig1}. 
We also investigated the effects of rotation on the axisymmetric modes ($m=0$) for various values of the vertical wave number $k_z$, and found that their growth rates are hardly affected, by less than one percent of $|v_{r,\rm sh}|/r_{\rm sh}$ in our study (Fig.~\ref{fig2}).
A global overview of the parameter space of the cylindrical SASI is displayed in Fig. \ref{fig3}, which indicates the azimuthal wavenumber $m$ of the most unstable mode for a wide range of shock radii and specific angular momentum. The asymmetric one-armed spiral mode is unstable in most of the parameter space, and always more unstable than without rotation. The growth rate of the mode $m=2$ exceeds that of the mode $m=1$ (by less than $10\%$ in the example of Fig.~1), as the specific angular momentum is increased.

\section{Discussion\label{discussion}}

\subsection{Instability mechanism} \label{mechanism}

As underlined in Sect.~3, the dynamical effect of the centrifugal force on the stationary flow is modest. We anticipated in Sect.~2 that the only linear effect of angular momentum is a Doppler shift of the eigenfrequency $\omega'=\omega-m\Omega(r)$, where $\Omega(r)$ is the local rotation frequency. This leaves the axisymmetric mode $m=0$ unaffected and explains the relative insensitivity of its growth rate with respect to the rotation rate, at least for moderate angular momentum. The strong effect of rotation on the growth rate of the spiral modes can thus be traced back to this Doppler shifted frequency. What is the mechanism of the instability ? As seen in the previous section, the destabilizing role of rotation does not seem related to the presence or absence of a corotation radius, thus discarding a Papaloizou-Pringle mechanism \citep{gol85}. Two possibilities have been proposed for the mechanism of SASI without rotation; one is the advective-acoustic mechanism \citep{fog00,fog01,fog02} and the other is the purely acoustic mechanism \citep{blo06}. Up to now, there is no satisfactory direct argument for the mechanism of the modes with a long wavelength. FGSJ07 used a WKB approximation to prove that the instability of the modes with a short wavelength is due to an advective-acoustic mechanism
and extrapolated this conclusion to the modes with a long wavelength, which are the most unstable. This method, recalled in Appendix~B, is based on the identification of acoustic waves and advected waves at a radius immediately below the shock surface, and the measurement of their coupling coefficients, above this radius due to the shock, and below this radius due to the flow gradients. These coupling processes are responsible for the existence of two cycles, namely a purely acoustic cycle characterized by an efficiency ${\cal R}$, and an advective-acoustic cycle characterized by an efficiency
${\cal Q}$.

By using the same method, the present study does not address directly the instability mechanism of long wavelength modes. However, the WKB approximation enables us to describe, in a conclusive manner, the instability mechanism of short wavelength modes affected by rotation.
First we checked that when the shock distance is increased ($r_{\rm sh}=20r_*$), the overtones are also unstable and their growth rate is an oscillatory function of the frequency similar to Fig.~7 of FGSJ07. The effect of rotation on the advective-acoustic cycle is illustrated by Fig~3, for the spiral modes $m=\pm1$ corresponding to the $10$-th overtone, as a function of the rotation rate. The cycle efficiency ${\cal Q}$ is strongly amplified by rotation if $m>0$,  while strongly damped if $m<0$. The stabilization of the counter-rotating spiral coincides with a marginally stable cycle ${\cal Q}\sim1$. The calculation of the amplification factor ${\cal R}$ of perturbations during each purely acoustic cycle indicates its stability (${\cal R}<1$). Contrary to the expectation of \citet{lam07} (see next subsection), rotation clearly favours the spiral mode of the advective-acoustic cycle. 

This consequence of rotation established unambiguously for short wavelength perturbations is identical to the influence of rotation on the fundamental mode of SASI: we consider this a new hint that the advective-acoustic mechanism can be extrapolated to low frequencies. The detailed analysis of the consequences of the Doppler shifted frequency on the increase of the advective-acoustic efficiency ${\cal Q}$ will be presented elsewhere (\cite{yam08}, in preparation).

\subsection{Comments on the Results of Laming (2007)} \label{comment}

The effect of rotation on the growth rate of SASI, established in Sect. 3 in a cylindrical geometry, is qualitatively similar to the effect conjectured by \cite{lam07} (hereafter L07). 
Nevertheless, their investigation about the instability mechanism led them to a different interpretation of the roles of the acoustic and advective-acoustic cycles.

We must point out a fundamental difference between the method of L07 and ours: by using a WKB approximation, we have carefuly defined the range of validity  of our method, namely short wavelength modes. This guarantees that the advective-acoustic interpretation of the instability mechanism is physical and robust, at least in some parameter range. In contrast, the existence of a purely acoustic instability is still a conjecture because the domain of validity of the method used by L07 is ambiguous: their analytical derivation of a dispersion relation when advection is included requires to neglect terms of order $(v_r /\omega r)$ while terms of order ${\cal M}$ are retained. This approximation is not supported by the results of their Fig.~2, which indicates that $(v_{r,{\rm sh}}/\omega r_{\rm sh})$ is comparable to or larger than ${\cal M}_{\rm sh}$ for the modes $l=0$ and $l=1$. An accurate description of this acoustic mode, even in a simplified set up, would be useful to gain confidence in its possible existence.

In addition to the question of the validity of the approximations used by L07, we find that our results invalidate their reasoning concerning the instability mechanism. They proposed that the advective-acoustic mechanism would be essential if $r_{\rm sh}/r_*\ge 10$, whereas a purely acoustic unstable process would be dominant for small shock radii, and they argued that rotation is a key ingredient to discriminate between the two mechanisms. When rotation is included, its effect on SASI has been attributed by L07 to a purely acoustic mechanism, despite the results of their Table 3. However, their view that rotation cannot possibly enhance the growth of the advective-acoustic cycle is clearly incorrect, at least for the short wavelength modes (our Fig.~\ref{fig4}). 

\subsection{Consequences of rotation on supernova explosions} \label{consequence}

The perturbative study of a simple cylindrical configuration has enabled us to cover a large parameter space of shock radii and rotation rates, in order to (i) demonstrate the linear selection of non-axisymmetric modes, (ii) establish a correlation between the preferred direction of the spiral SASI and the rotation of the collapsing core, (iii) identify the advective-acoustic mechanism at work for short wavelength spiral perturbations.

The fact that rotation favours a spiral mode $m=1,2$ in a cylindrical flow seems directly connected to the property observed by \cite{blo07a} in their 3D simulations including rotation. Tracing back the main influence of rotation to the local Doppler shifted frequency $\omega-m\Omega$, we may indeed expect a similar destabilization of the spiral modes with a positive value of $m$, a stabilization of the counter-rotating ones, and a comparatively weak influence on the axisymmetric modes. 

Even a moderate amount of angular momentum results in a shortening of the growth time of SASI through the destabilization of a non-axisymmetric mode. The promising consequences of SASI on both the explosion mechanisms and the pulsar kick could thus be considerably modified, since they were established on the basis of axisymmetric numerical simulations \citep{bur06,bur07,mar07,sch04,sch06}.
Our study suggests that the effect of rotation on the linear phase of SASI can be safely neglected only for slowly rotating progenitors with a specific angular momentum $L\ll 2\pi \cdot 10^{14}$ cm$^2/$s. Although a fast growth of SASI might be helpful to an early shock revival, the dynamical effects of a spiral mode $m=1$, and even $m=2$, on the possible explosion mechanisms are not known yet. 

If the direction of the kick were determined by the geometry of the most unstable $l=1$ SASI mode, our perturbative approach would suggest a kick-spin misalignment. The strength of the equatorial kick may be diminished by the domination of a symmetric mode $m=2$. It is worth noting however that the relationship between the timescale of the most unstable SASI mode and the onset of explosion is not straightforward, and should be evaluated by future 3D numerical simulations. Our linear approach modestly aims at guiding our intuition for the interpretation of these simulations.


\appendix

\section{Derivation of the Basic Equations}

\subsection{Basic Equations}

The basic equations describing the flow are
\begin{eqnarray}
\frac{\partial \rho}{\partial t}+\nabla\cdot(\rho\mbox{\boldmath $v$})=0,\label{cont}\\
\frac{\partial \mbox{\boldmath $v$}}{\partial t}+\mbox{\boldmath $w$}\times \mbox{\boldmath $v$}
+\nabla \left(\frac{|\mbox{\boldmath $v$}|^2}{2}+\frac{c^2}{\gamma-1}+\Phi \right)=\frac{c^2}{\gamma} \nabla S,\label{mom}\\
\frac{\partial S}{\partial t}+\mbox{\boldmath $v$}\cdot \nabla S=\frac{\cal L}{P}.\label{energy}
\end{eqnarray}
Small amplitude perturbations are superimposed onto the above equations.
From the transverse components $(\theta,z)$ of the Euler equation, the vorticity can be expressed as follows:
\begin{eqnarray}
\delta w_\theta&=&{ik_z\over v_r}\left(\delta f-{L\over r}\delta v_\theta+\delta q-{c^2\over\gamma}\delta S\right)-{i\omega'\over v_r}\delta v_z,\label{wthet}\\
\delta w_z&=&i\omega{\delta v_\theta\over v_r}+{im\over rv_r}\left({c^2\over\gamma}\delta S-\delta f-\delta q\right).\label{wzed}
\end{eqnarray}
From the definition of the  vorticity vector,
\begin{eqnarray}
\delta w_r&\equiv&{im\over r}\delta v_z-ik_z\delta v_\theta,\\
{{\rm d} r\delta v_\theta\over{\rm d} r}&=&im\delta v_r+r\delta w_z,\label{drvt}\\
{{\rm d} \delta v_z\over{\rm d} r}&=&ik_z\delta v_r-\delta w_\theta.\label{dvz}
\end{eqnarray}
Using the two equations (\ref{wthet}) and (\ref{wzed}) and the definition of vorticity leads to:
\begin{eqnarray}
\left({{\rm d}\over{\rm d} r}-{i\omega'\over v_r}\right)(r\delta w_r)&=&0.\label{conswr}
\end{eqnarray}
The radial Euler equation combined with Eq.~(\ref{eq8}) and (\ref{eq9}) is
\begin{eqnarray}
{{\rm d} \delta f\over {\rm d} r}&=&i\omega\delta v_r + {L\over r}\delta w_z+{i\omega'\over v_r}
\left({c^2\over\gamma}\delta S-\delta q\right).\label{dfdr}
\end{eqnarray}
Guided by the conservation of $\delta K$ in a radial flow \citep{fog01}, let us define the quantities $\delta K_1,\delta K_2$ as follows:
\begin{eqnarray}
\delta K_1&\equiv&v_r r\delta w_z-im\left({c^2\over\gamma}\delta S-\delta q\right),\label{defk1}\\
&=&i\omega r\delta v_\theta-im\delta f,\label{dk1}\\
\delta K_2&\equiv&v_r\delta w_\theta+ik_z\left({c^2\over\gamma}\delta S-\delta q\right),\\
&=&-i\omega' \delta v_z+ik_z\left(\delta f-{L\over r}\delta v_\theta\right),\label{dk2}
\end{eqnarray}
where Eq.~(\ref{dk1}) and Eq.~(\ref{dk2}) are deduced from Eqs.~(\ref{wthet}) and (\ref{wzed}).
The flow quantities $\delta v_\theta$, $\delta v_z$, $\delta w_\theta$, $\delta w_z$ can be expressed with $\delta K_1$ and $\delta K_2$ using Eqs.~(\ref{defk1}-\ref{dk2}):
\begin{eqnarray}
\delta v_\theta &=& {m\over r}{\delta f\over\omega}-{i\delta K_1\over \omega r},\label{dvtheta}\\
\delta v_z &=& k_z{\delta f\over\omega}+{i\over\omega'}\left(\delta K_2+{Lk_z\over\omega r^2}\delta K_1\right),\\
\delta w_\theta&=&{ik_z\over v_r}\left(\delta q -{c^2\over \gamma}\delta S\right)+{\delta K_2\over v_r},\\
\delta w_z&=&-{im\over v_r r}\left(\delta q -{c^2\over \gamma}\delta S\right)+{\delta K_1\over rv_r}.\label{dwz}
\end{eqnarray}
Using Eq.~(\ref{drvt}), (\ref{dvz}) and (\ref{dfdr}), we can prove that 
\begin{eqnarray}
\left({{\rm d}\over{\rm d} r}-{i\omega'\over v_r}\right)\delta K_1&=&0,\label{consk1}\\
\left({{\rm d}\over{\rm d} r}-{i\omega'\over v_r}\right)\delta K_2&=&-{2L\over r^2}\delta w_r.\label{consk2}
\end{eqnarray}
The perturbations of radial velocity, sound speed and density are related to $f,h,\delta S,\delta q$ as follows:
\begin{eqnarray}
{\delta v_r\over v_r}&=&{1\over 1-{\cal M}^2}\left(\delta h+\delta S-{\delta f\over c^2}+{L\over rc^2}\delta v_\theta
-{\delta q\over c^2}\right),\\
{\delta c^2\over c^2}&=&{\gamma-1\over 1-{\cal M}^2}\left({\delta f\over c^2}-{L\over rc^2}\delta v_\theta
+{\delta q\over c^2}-{\cal M}^2 \delta h-{\cal M}^2\delta S\right),\\
{\delta \rho\over \rho}&=&{1\over 1-{\cal M}^2}\left({\delta f\over c^2}-{L\over rc^2}\delta v_\theta+{\delta q\over c^2}
-{\cal M}^2 \delta h-\delta S\right).
\end{eqnarray}
The continuity of mass flux is
\begin{eqnarray}
{{\rm d} \delta h\over{\rm d} r}={i\omega'\over v_r}{\delta\rho\over \rho}-{im\over r v_r}\delta v_\theta.
\end{eqnarray}
The differential system satisfied by $\delta f, \delta h,\delta S, \delta q$ is
\begin{eqnarray}
{{\rm d} \delta f\over{\rm d} r}&=&{i\omega c^2\over v_r(1-{\cal M}^2)}\left\lbrace
{\cal M}^2 \delta h -{\cal M}^2{\omega'\over\omega}{\delta f\over c^2}
+\left\lbrack1+(\gamma-1){\cal M}^2\right\rbrack{\delta S\over\gamma}-{\delta q\over c^2}
 \right\rbrace\nonumber\\
 &&+{L\over v_r r^2}{\delta K_1\over 1-{\cal M}^2},\label{appdf}\\
{{\rm d} \delta h\over{\rm d} r}&=&{i\omega'\over v_r(1-{\cal M}^2)}\left(
{\mu^{2} \over c^{2}}{\omega'\over \omega} \delta f
 -{\cal M}^2 \delta h- \delta S+{\delta q\over c^2}\right)\nonumber\\
 &&-{\delta K_1\over\omega r^2 v_r}
 \left( m+{L\omega'\over c^2(1-{\cal M}^2)}\right)
 \label{dhp},\\
{{\rm d} \delta S\over{\rm d} r}&=&{i\omega'\over v_r}\delta S
+\delta\left({{\cal L}\over Pv_r}\right), \label{dsp}
\\
{{\rm d} \delta q\over {\rm d} r}&=&{i\omega'\over v_r}\delta q
+\delta\left({{\cal L}\over \rho v_r}\right). \label{difq}
\end{eqnarray}
The transverse velocity perturbations $\delta v_\theta,\delta v_z$ at the shock are expressed in Appendix A.2. by:
\begin{eqnarray}
\delta v_{\theta,{\rm sh}}&=&{im\over r_{\rm sh}}\Delta\zeta(v_{r,1}-v_{r,{\rm sh}}),\label{theta}\\
\delta v_{z,{\rm sh}}&=&ik_z\Delta\zeta(v_{r,1}-v_{r,{\rm sh}}). \label{vertical}
\end{eqnarray}
Together with the boundary conditions (Eqs.~(\ref{eq12}-\ref{eq15})) established in Appendix A.2., we deduce from the definition of $\delta w_r,\delta K_1,\delta K_2$ that these three quantities vanish at the shock. From the conservation Eqs.~(\ref{conswr}), (\ref{consk1}), (\ref{consk2}), we conclude that $\delta w_r$, $\delta K_1$ and $\delta K_2$ are uniformly zero throughout the flow. The flow quantities $\delta v_\theta$, $\delta v_z$, $\delta w_\theta$ expressed in Eqs.~(\ref{dvtheta}-\ref{dwz}) are thus simplified accordingly, and the differential system (\ref{appdf}-\ref{difq}) is transformed into the simpler Eqs.~(\ref{eq6}-\ref{eq9}). 

\subsection{Boundary Conditions}

The Rankine-Hugoniot relation is written as

\begin{eqnarray}
\rho_1({\bf v}_1 -{\bf v}_{\rm s})\cdot \frac{\bf n}{|{\bf n}|}&=&\rho_{\rm sh}({\bf v}_{\rm sh} -{\bf v}_{\rm s})\cdot \frac{\bf n}{|{\bf n}|},
\label{b1}\\
\rho_1\left\{({\bf v}_1 -{\bf v}_{\rm s})\cdot \frac{\bf n}{|{\bf n}|}\right\}^2+\frac{\rho_1 c_1^2}{\gamma}&=&\rho_{\rm sh}\left\{({\bf v}_{\rm sh} -{\bf v}_{\rm s})\cdot \frac{\bf n}{|{\bf n}|}\right\}^2+\frac{\rho_{\rm sh} c_{\rm sh}^2}{\gamma},
\label{b2}\\
({\bf v}_1 -{\bf v}_{\rm s})\cdot {\bf t_1}&=&({\bf v}_{\rm sh} -{\bf v}_{\rm s})\cdot {\bf t_1},
\label{b3}\\
({\bf v}_1 -{\bf v}_{\rm s})\cdot {\bf t_2}&=&({\bf v}_{\rm sh} -{\bf v}_{\rm s})\cdot {\bf t_2},
\label{b4}\\
\frac 12 \left\{({\bf v}_1 -{\bf v}_{\rm s})\cdot \frac{\bf n}{|{\bf n}|}\right\}^2 +\frac{c_1^2}{\gamma-1}&=&\frac 12 
\left\{({\bf v}_{\rm sh} -{\bf v}_{\rm s})\cdot \frac{\bf n}{|{\bf n}|}\right\}^2
+\frac{c_{\rm sh}^2}{\gamma-1},
\label{b5}
\end{eqnarray}
where ${\bf v}_{\rm s}$ is the velocity vector of the shock surface and ${\bf n}$, ${\bf t_1}$ and ${\bf t_2}$ are the vector normal and tangent to the shock surface which is written at first order as follows,

\begin{eqnarray}
{\bf n}&=&\left(1,-\frac{1}{r_{\rm s}}\frac{\partial r_{\rm s}}{\partial \theta},
-\frac{\partial r_{\rm s}}{\partial z}\right),\\
{\bf t_1}&=&\left(\frac{\partial r_{\rm s}}{\partial \theta},
r_{\rm s},0\right),\\
{\bf t_2}&=&\left(\frac{\partial r_{\rm s}}{\partial z},0,1\right).
\end{eqnarray}
Considering small perturbations of the above Eqs, (\ref{b1})-(\ref{b5}), we obtain 
\begin{eqnarray}
\rho_{\rm sh}v_{r,{\rm sh}}\delta h_{\rm sh}+i\omega' \Delta \zeta(\rho_{\rm sh}-\rho_1)=\Delta \zeta \left[\frac{d}{dr} (\rho v_r)_1 -\frac{d}{dr}(\rho v_r)_{\rm sh}\right] ,\\
v_{r,{\rm sh}}^2 \delta \rho_{\rm sh}+2\rho_{\rm sh}v_{r,{\rm sh}}\delta v_{r,{\rm sh}}+\frac{2}{\gamma}\rho_{\rm sh}c_{\rm sh}\delta c_{\rm sh}+\delta \rho_{\rm sh}\frac{c_{\rm sh}^2}{\gamma} \nonumber\\
=\Delta \zeta\left[\frac{d}{dr}(\rho v_r^2 +P)_1 -\frac{d}{dr}(\rho v_r^2 +P)_{\rm sh}\right],\\
\delta f_{\rm sh}-v_{{\theta},{\rm sh}}\delta v_{\theta,{\rm sh}}+\delta q_{\rm sh}+i\omega' \Delta \zeta(v_{\rm sh}-v_1) \nonumber\\
=\Delta \zeta \left[\frac{d}{dr}\left(\frac{v_r^2}{2}+\frac{c^2}{\gamma-1}\right)_1 -\frac{d}{dr}\left(\frac{v_r^2}{2}+\frac{c^2}{\gamma-1}\right)_{\rm sh}\right],
\end{eqnarray}
and Eqs. (\ref{theta})-(\ref{vertical}). Using the relations in the steady flow,
\begin{eqnarray}
\frac{d}{dr}(\rho v_r)&=&-\frac{\rho v_r}{r},\\
\frac{d}{dr}(\rho v_r^2 +P)&=&\rho\frac{d\Phi}{dr}-\frac{\rho v_r^2}{r},\\
\frac{d}{dr}\left(\frac{v_r^2 +v_{\theta}^2}{2}+\frac{c^2}{\gamma-1}\right)&=&\frac{\cal L}{\rho v_r}+\frac{d\Phi}{dr},
\end{eqnarray}
we obtain the boundary conditions (\ref{eq12})-(\ref{eq15}).

\section{WKB Method for the Calculation of the Amplification Coefficients}

\begin{figure}
\epsscale{1.0}
\plotone{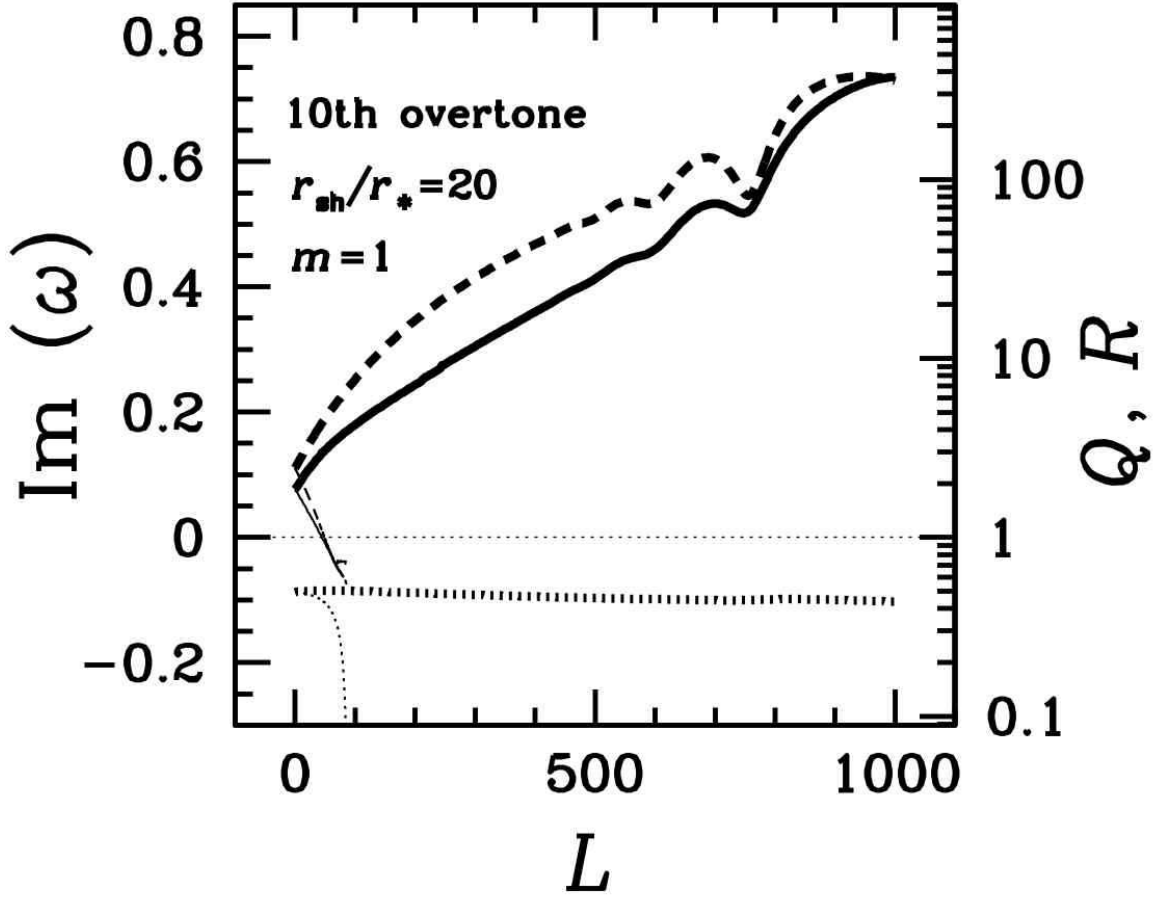}
\caption{In a flow with $r_{\rm sh}/r_*=20$, the two spiral components $m=+1$ (thick lines) and $m=-1$ (thin lines) of the 10-th overtone are analyzed as a function of the specific angular momentum $L$, by solving the boundary value problem (linear scale) and computing the cycle efficiencies ${\cal Q}$, ${\cal R}$ in the WKB approximation (logarithmic scale). The growth rates Im($\omega$) (full lines) are compared with the advective-acoustic efficiencies ${\cal Q}$ (dashed lines, logarithmic scale) and the purely acoustic efficiencies ${\cal R}$ (dotted lines). Units are identical to Fig.~1. 
\label{fig4}}
\end{figure}

In order to interpret the complex eigenfrequency $\omega$ of a given eigenmode, we compute the efficiency of the advective-acoustic and purely acoustic cycles associated to the real frequency $\omega_r\equiv$ Re$(\omega)$ of this eigenmode.
The coefficient ${\cal R}(\omega_r)$ is defined by the amplification of perturbations after one purely acoustic cycle, initiated at the shock by an acoustic wave propagating downward. ${\cal Q}(\omega_r)$ and ${\cal Q}^q(\omega_r)$ measure the amplification of pertubations through an advective-acoustic cycle, initiated at the shock by the advection of an entropy perturbation $\delta S$ with $\delta q=0$, or a heat perturbations $\delta q$ with $\delta S=0$ respectively.
Each of these coefficients  ${\cal R}$, ${\cal Q}$ and ${\cal Q}^q$ is the product of the coupling coefficient at the shock (${\cal R}_{\rm sh}$, ${\cal Q}_{\rm sh}$, or ${\cal Q}_{\rm sh}^q$), multiplied by the coupling coefficient through the flow (${\cal R}_{\nabla}$, ${\cal Q}_{\nabla}$ or ${\cal Q}^q_{\nabla}$). The technique of calculation of each factor is the same as that described in the Appendix D of FGSJ07. The calculations are based on the decomposition of the variables onto the basis of advected and acoustic perturbations, which is exact when the flow is uniform. Even when the flow is moderately inhomogeneous, a similar decomposition is obtained using a WKB approximation. Since the definitions of the variables $\delta f,\delta h,\delta S,\delta q$ employed in this paper are slightly different from those in FGSJ07, the decomposition is modified as follows:
\begin{eqnarray}
\delta f=\delta f^{+}+\delta f^{-}+\delta f^{S}+\delta f^{q},\\
\delta h=\delta h^{+}+\delta h^{-}+\delta h^{S}+\delta h^{q}.
\end{eqnarray}
The superscripts $+$, $-$, $S$, $q$ refer to the contributions of the ingoing and outgoing acoustic wave, the advected quantities $\delta S$ and $\delta q$, respectively. Adopting the WKB approximation, the quantities $\delta f^{\pm},\delta h^{\pm}$ associated with the acoustic waves satisfy the differential system (\ref{eq6}-\ref{eq9}) where $\delta S=0$ and $\delta q=0$, and the radial derivatives are replaced by a multiplication by $ik_{\pm}$:
\begin{eqnarray}
ik_{\pm}\delta f^{\pm}=\frac{i\omega c^2}{v_r (1-{\cal M}^2)}\left({\cal M}^2 \delta h^{\pm}
-{\cal M}^2 \frac{\omega'}{\omega}\frac{\delta f^{\pm}}{c^2}\right),\\
ik_{\pm}\delta h^{\pm}=\frac{i\omega'}{v_r (1-{\cal M}^2)}\left(\frac{\mu^2}{c^2}\frac{\omega'}{\omega}\delta f^{\pm}-{\cal M}^2 \delta h^{\pm}\right).
\end{eqnarray}
The dispersion relation of acoustic waves corresponds to:
\begin{eqnarray}
k_{\pm}=\frac{\omega'}{c}\frac{{\cal M}\mp \mu}{1-{\cal M}^2}.
\end{eqnarray}
The advected quantities satisfy the differential system (\ref{eq6}-\ref{eq9}) where the radial derivatives are replaced by a multiplication by $ik_0$, where $k_0\equiv \omega'/v$.
\begin{eqnarray}
ik_0 \delta f^S=\frac{i\omega c^2}{v_r(1-{\cal M}^2)}\left\{{\cal M}^2 \delta h^S
-{\cal M}^2 \frac{\omega'}{\omega}\frac{\delta f^S}{c^2}
+[1+(\gamma -1){\cal M}^2 ]\frac{\delta S}{\gamma}\right\},\\
ik_0\delta h^S=\frac{i\omega'}{v_r(1-{\cal M}^2)}\left(\frac{\mu^2}{c^2}\frac{\omega'}{\omega}\delta f^S-{\cal M}^2 \delta h^S-\delta S\right),\\
ik_0 \delta f^q=\frac{i\omega c^2}{v_r(1-{\cal M}^2)}\left({\cal M}^2 \delta h^q
-{\cal M}^2 \frac{\omega'}{\omega}\frac{\delta f^q}{c^2}
-\frac{\delta q}{c^2}\right),\\
ik_0\delta h^q=\frac{i\omega'}{v_r(1-{\cal M}^2)}\left(\frac{\mu^2}{c^2}\frac{\omega'}{\omega}\delta f^q-{\cal M}^2 \delta h^q+\frac{\delta q}{c^2}\right).
\end{eqnarray}
Solving these two sets of equations leads to:
\begin{eqnarray}
\delta h^{\pm}={\pm}\frac{\omega'}{\omega}\frac{\mu}{{\cal M}c^2}\delta f^{\pm},\\
\delta f^{S}=\frac{\omega}{\omega'}\frac{1-{\cal M}^2}{1-\mu^2 {\cal M}^2}\frac{c^2}{\gamma}\delta S,\\
\delta h^{S}=\frac{\omega'}{\omega}\frac{\mu^2}{c^2}\delta f^S -\delta S,\\
\delta f^{q}=-\frac{\omega}{\omega'}\frac{1-{\cal M}^2}{1-\mu^2 {\cal M}^2}\delta q,\\
\delta h^{q}=\frac{1-\mu^2}{1-\mu^2 {\cal M}^2}\frac{\delta q}{c^2}.
\end{eqnarray}
The coupling coefficients ${\cal R}_{\rm sh}$, ${\cal Q}_{\rm sh}$ and ${\cal Q}^q_{\rm sh}$ are obtained by decomposing the variables at the boundary described by Eqs.~(\ref{eq12}-\ref{eq15}) onto the basis of acoustic and advected perturbations, immediately below the shock:
\begin{eqnarray}
\delta f_{\rm sh}&=&  \delta f_{\rm sh}^++\delta f_{\rm sh}^-+\delta f_{\rm sh}^S+\delta f_{\rm sh}^q,\\
{\cal R}_{\rm sh}&\equiv& {\delta f_{\rm sh}^+\over \delta f_{\rm sh}^-},\\
{\cal Q}_{\rm sh}&\equiv& {\delta f_{\rm sh}^S\over \delta f_{\rm sh}^-},\\
{\cal Q}^q_{\rm sh}&\equiv& {\delta f_{\rm sh}^q\over \delta f_{\rm sh}^-}.
\end{eqnarray}
The three coupling coefficients ${\cal R}_{\nabla}$, ${\cal Q}_{\nabla}$ and ${\cal Q}^q_{\nabla}$ are calculated by measuring numerically,  at a radius $R$ immediately below the shock ($R= r_{\rm sh}$), the acoustic feedback $\delta f^-(R)$ that would be produced, either by an ingoing purely acoustic perturbation $\delta f^+$, or a purely advective perturbation $\delta f^S$, or $\delta f^q$. Each of these three coefficients is calculated by integrating the differential system (\ref{eq6}-\ref{eq9}) from the radius $R$ down to the accretor surface. For example, the boundary condition used at $r=R$ for the calculation of ${\cal Q}_{\nabla}$ involves a perturbation of entropy and vorticity $\delta f^S(R)$, and the right amount of acoustic feedback $\delta f^-(R)$, 
\begin{eqnarray}
\delta f(R)&=&\delta f^S(R)+\delta f^-(R),
\end{eqnarray}
such that the inner boundary condition at the accretor surface is satisfied.
The coupling coefficient ${\cal Q}_\nabla$ measures the efficiency of this acoustic feedback:
\begin{eqnarray}
{\cal Q}_\nabla&\equiv&{\delta f^-(R)\over \delta f^S(R)}.
\end{eqnarray}
Since ${\cal Q}^q$ is negligible compared to both ${\cal Q}$ and ${\cal R}$, we discuss only ${\cal R}$ and ${\cal Q}$ in the text.
The WKB decomposition is a good approximation when the inhomogeneity caused by the convergence of the flow, gravity and cooling is moderate within a wavelength of the perturbation.
Since we use this decomposition immediately below the shock front, the approximation is valid when the inhomogeneity of the steady flow just below the shock is sufficiently small. The amplification coefficients ${\cal Q}$ and ${\cal R}$ illustrated in our Fig.~\ref{fig4} were computed in a flow with a large shock radius ($r_{\rm sh}=20r_{\ast}$), for a short wavelength mode (tenth overtone), in order to obtain reliable results.

\clearpage

\end{document}